%% file: main.tex
\documentclass[11pt]{article}
\usepackage{graphicx}
\usepackage[margin=1.25in]{geometry}
\usepackage[usenames,dvipsnames]{color}
\usepackage{url}
\usepackage[colorlinks = true,
            linkcolor = blue,
            urlcolor  = blue,
            citecolor = blue,
            anchorcolor = blue]{hyperref}
\usepackage{authblk}

\input workshopsymbols.tex

\newcommand\snowmass{\begin{center}\rule[-0.2in]{\hsize}{0.01in}\\\rule{\hsize}{0.01in}\\
\vskip 0.1in Submitted to the  Proceedings of the US Community Study\\ 
on the Future of Particle Physics (Snowmass 2021)\\ 
\rule{\hsize}{0.01in}\\\rule[+0.2in]{\hsize}{0.01in} \end{center}}

\author[1]{Matt Bellis}
\author[2]{Bhubanjyoti Bhattacharya}
\author[3]{David DeMuth}
\author[4]{Julie Hogan}
\author[5]{Kathrine Laureto}
\author[6]{Sudhir Malik\thanks{sudhir.malik@upr.edu}}
\author[7]{Ben Pearson}
\affil[1]{\small Department of Physics \& Astronomy, Siena College, Loudonville NY 12211, USA} %515 Loudon Road, Loudonville, NY 12211-1462, United States}
\affil[2]{Department of Natural Sciences, Lawrence Technological University, Southfield MI, 48075, USA}% 48075, USA}Department of Natural Sciences, 
\affil[3]{Department of Science, Valley City State University, Valley City ND, 58072, USA} %58072}Science Department, 
\affil[4]{Department of Physics \& Engineering, Bethel University, St.~Paul MN, 55112, USA} %3900 Bethel Dr, St Paul, MN 55112, USA}Department of Physics and Engineering, 
\affil[5]{Accelerator Division, Fermi National Accelerator Laboratory, Batavia IL, 60510, USA} % 60510, USA}Accelerator Division, 
\affil[6]{Department of Physics, University of Puerto Rico Mayagüez, Mayagüez PR, 00682, USA}% 00682, USA}Physics Department, 
\affil[7]{Department of Engineering \& Physics, Northwest Nazarene University, Nampa ID, 83686, USA}%623 S University Blvd, Nampa, ID 83686, USA}

\title{Enhancing HEP research in predominantly undergraduate institutions and community colleges}

\begin{document}
\maketitle

%\pubblock

%\medskip

 \begin{abstract}
     The long-term success of HEP lies in expanding inclusiveness beyond national labs and academic research institutions to a vast community of predominantly undergraduate institutions (PUI) and community colleges (CC). Institutions such as PUIs and CCs offer an early starting point in the pipeline that can mitigate issues of lack of diversity and underrepresented participation of different groups in HEP. However, there are many underlying systemic, structural, and cultural challenges that need to be addressed collectively. Experimental collaborations are largely populated by national labs and research-focused academic institutions (non-PUIs). The faculty at PUIs and CCs have a high teaching load that is detrimental to their research participation. In addition, there is a lack of guidance, access, and tough competition for securing research funding. The students also suffer from a lack of research infrastructure and technical equipment that can only be found at national labs and larger universities. There are existing successful efforts to enhance the HEP research experience of students and faculty members. This paper discusses ways to leverage these to provide more research opportunities and establish a sustainable national program targeting specifically the issues faced by communities at PUIs and CCs. The need for research mentoring and skill building for faculty members is also laid out. The changes discussed in this paper would make a direct impact on the current spectrum of challenges.

\end{abstract}

\newpage
\snowmass

\def\thefootnote{\fnsymbol{footnote}}
\setcounter{footnote}{0}

\section{Introduction}

The spectrum of challenges for doing HEP research at PUIs\cite{Ref1} are common to all areas of research, though the details might differ. A lot has been already researched and documented about these problems~\cite{Ref1b1,Ref2,Ref3,Ref4,Ref5,Ref6,Ref7}. The field of HEP worldwide has been led by the National Labs and non-PUIs, complemented by participation from universities and labs of similar nature. On the experimental side the need for expensive state-of-the-art facilities like particle accelerators and complex particle detectors leads to a highly collaborative nature of the field that makes it unique and presents several challenges. The funding for basic sciences has been plateauing for over two decades. In parallel, the influx of young scientists outnumbers those who retire from the system, making fewer resources and jobs available.

Many young postdocs and graduating Ph.D.~students leave the field after investing years of time into HEP research, as academic opportunities at research-focused institutes are too few and therefore highly competitive. An interesting survey by the American Institute of Physics reveals that lack of funding and difficulty obtaining employment are significant factors that influence career choices made by physicists~\cite{Ref8}. A small number of those driven to both teach and research, however, find employment at PUIs as an attractive academic career. However, as mentioned before, this comes with specific challenges including limited time for research due to higher teaching loads, lack of funding for research (whether start-up funds or federal grants), and an obvious lack of Ph.D.~students who are the backbone of committed research programs. In the case of HEP this is even more accentuated, as there are very few PUIs participating in HEP experiments. In this paper we discuss these challenges and the motivation to mitigate them, with a focus on HEP. It should be noted that these challenges are not unique to HEP and mirror other scientific fields 
%such as astrophysics and nuclear physics. While the authors of this paper are themselves HEP scientists, we note that our recommendations map on to other scientific fields 
with researchers at PUIs. 

\section{Benefits of performing research at PUIs and CCs}

It is natural to question whether or not cutting-edge research can be done at a PUI or CC, given
the high teaching load and the expertise disparity between undergrads and grad students. 
With finite resources, does it make sense to funnel some of those
resources to PUIs? Given the collaborative nature of HEP experiments, as well as the HEP
theoretical community, it is difficult
to pinpoint whether or not any one person or one institution is indispensable to the end
goal of discovering New Physics and so we choose not to focus on that. Instead, we make a strong argument that PUIs and CCs engage a {\it fundamentally different} student population and often
provide a very different set of experiences for the students when compared to larger research
institutions. It is important for any scientific field to reflect the broader society in which it
operates, otherwise it is in danger of being disconnected from
that society, leading to a lack of support. 

There are a number of reasons why a student might attend a PUI or CC. The CCs tend not to offer
a residential experience and have a lower cost per credit hour than other colleges. This attracts
students who are looking to save money, perhaps because they are in a lower economic bracket 
or because they are unsure of exactly what they want to do at a 4-year institution and need 
the time to figure it out. CC students are also often holding down jobs while they take classes. 
These non-traditional students can still be just as passionate about science as their 
peers at other colleges and universities and should have access to teachers who are regularly 
engaged with Big Science, writ large. 

Students who choose a 4-year liberal arts college, may do so for similar economic reasons, but 
are also often drawn to the small class size and the advertised closer connection with their 
professors. They can be overwhelmed by the large classes in many research and state universities
or may need to stay close to home for any number of reasons. Again, these students can be incredibly
passionate about science and are at a ``pivot-point" in their lives as they decide what they want
to do after college. Do they get a job? Pursue grad school? Both? Connection with professors who 
are actually {\it doing} the type of research they are considering can help them make the best
informed decision. Having the opportunity to work on this type of research can be life-changing
for any undergrad. But if these opportunities only take place at larger research institutions, 
we are potentially excluding, or at least minimizing, a particular cross section of
individuals. 

Across many STEM fields, there is an interest in plugging the ``leaky pipeline", whether that is women
or POC (persons of color) or students with disabilities or other URMs (under-represented minorities).
There is no one approach that will solve this problem, but it does mean that we need to provide
students with positive experiences, both in and outside of the classroom, such that
they develop the confidence that they can succeed in that field and that they {\it want} to succeed
in that field. 
To that end, we want expose as many students as possible to the opportunities in HEP work as early
as possible in their education. 
PUIs provide a prime opportunity to offer research experiences that can mitigate these leaky pipelines.
Faculty at PUIs are particularly encouraged to reach out to individual students in an informed manner
and are often trained to engage with DEI issues. Because their time is not divided with supervising
responsibilities to graduate students, they have opportunities and incentives to nurture URM students
who might need that extra encouragement. 

It is also important to recognize that faculty at CCs and PUIs are often encouraged to work with
or teach to students who are not interested in a career in HEP research or even in a STEM field. 
In some cases, faculty at these schools, which often have smaller pools of majors from which
to draw, will have research students from the Engineering or Computer Science departments, 
particularly if they have funding. The students that work with these faculty have amazing
experiences that they take to their subsequent non-HEP careers. 

Many liberal arts colleges have significant ``{\it core curriculum}" or 
``{\it general education}" requirements.
This means that science students are required to take some number of humanities courses
and humanities or business majors are required to take some number of science or quantitative
courses. This is a {\it fantastic} opportunity to convey the excitement of and wonder of our
work to non-majors, almost all of whom will someday be voters who will weigh in on how taxes
are spent and what projects should be funded. One of the authors (Bellis) teaches a course
called {\it ``Quarks, Quanta, and Quasars"}, aimed at non-science majors. 60+ students
have gone through the course in the 3 terms it has been run, all of whom got to learn about 
developments in quantum computing, dark matter and dark energy, the Large Hadron Collider,
and many more topics. This material can, of course, be taught by faculty who are not 
actively engaged in research, but the material hits different (as the kids say) if the instructor
has a personal connection to the work. 

It should also be pointed out that because of the small class-sizes and the makeup of the
student body, faculty at PUIs and CCs provide a different perspective on teaching and mentoring
to their collaborators at larger institutions. They are often either required or strongly 
encouraged to engage more in DEI workshops, LGBT+ and Ally training, or guidance in dealing with
mental health issues. This cross-pollination of ideas can take place either in informal settings
or more formal collaboration projects. It is perhaps not a coincidence that two of the authors 
on this whitepaper, faculty at PUIs, (Bellis, Hogan) are co-conveners of the CMS MREFC 
(Major Research Equipment and Facilities Construction) Underrepresented Minority Internship 
program and that one of them (Hogan) is also a co-convener of the CMS
Mentorship program. Related, because most CCs and PUIs emphasize a commitment to teaching in the
classroom, these faculty can bring their experiences with education research to the broader
collaboration's training opportunities, such as the CMS Data Analysis School.

The diversity of the student body and student experience seem to suggest that there is a real
value-added in engaging faculty at PUIs and CCs in HEP work, both theory and experiment. However,
there are challenges for the faculty at these institutions. Later sections of this paper explore
those challenges as well as opportunities to meet those challenges. 

\section{Challenges for research at PUIs and CCs}
The challenges of establishing a successful research program at a PUI fall along three general fronts: research funding, participation in large collaborations, and the institutional culture of a PUI. 

\subsection{Research funding}

The most significant and direct challenge to the ability to perform research in HEP while being a faculty at PUI is acquiring research funding. The nature of HEP research can be experimental or theoretical and each has its own flavor of hurdles. The nature of experimental HEP is collaborative, driven by the need for expensive accelerator facilities, complex detector apparatuses, and extensive computing infrastructure. Participation in HEP experiments requires funding for certain essential activities like travel to the experimental site, wages for summer research, and annual contributions to the operation and maintenance of the experiment. At some PUIs funding is also required for the faculty member to have dedicated time available to perform research and pursue professional development, such as attending workshops and conferences. Funding for postdocs and students is a huge asset to a research program, and is often indispensable. Buying equipment like laptops for the research group and a small lab setup might also be a requirement. Support for almost all of these cannot be found at a typical PUI, where start-up funding is rare. These factors necessitate writing grants to federal agencies which are most likely the only source of HEP research funding.

For HEP there are funding opportunities from NSF~\cite{Ref9} called ``Facilitating Research at Primarily Undergraduate Institutions" that are very critical for young faculty members at PUIs. This funding opportunity requires additional writing in the form of an impact statement, and the proposal is subjected to the same rigorous peer review~\cite{Ref10} as a regular research grant. This implies that PUI faculty members are competing for funds from the same source as their non-PUI peers. Since experience shows that postdoc support is rare for faculty at PUIs, funding proposals are typically written individually by the faculty member. Faculty at PUIs also have a much higher teaching load compared to colleagues at non-PUIs. All of these issues makes the process of acquiring grants much more challenging for faculty at PUIs compared to those at non-PUIs who are required to do less writing and typically prepare proposals in large groups.

In general, the NSF programs such as Elementary Particle Physics (theory and experiment) do not have a history of supporting course buyouts, leaving it to the faculty member to negotiate a teaching load that still allows time for research. This is usually not in the form of a {\it reduced} course load, but perhaps consideration in avoiding preparing for multiple new courses. 

\subsection{Large experiment participation}

A typical HEP collaboration has hundreds of international collaborators from several universities. These universities represent the best institutes from respective countries. It is a highly competitive environment within a collaboration to make your mark, be it detector work or physics analysis. Since every project within the experimental collaboration involves multiple institutes it is very important to make sure that one's individual contribution is clearly acknowledged. An experiment does not care much for the teaching load of a collaborating faculty and this is what puts PUI faculties at a great disadvantage. It is very hard to compete and contribute individually at the level of collaborators from R1 universities who typically teach 2 courses or less per year. Scientists from labs and research institutes do not teach by design of their full time research jobs. Lab scientists also rarely compete for grants with their university faculty counterparts. At HEP experiment one is expected to do a few months of service work (like data taking, detector maintenance, upgrade activities and computing operations). In addition one has to do physics analysis which typically take 2-3 years from start to publish. With high teaching load, lack of postdoc and grad student support, a PUI faculty can barely keep up with research activity. Traveling to experimental site is possible only during university holidays but needs grant support. Though remote technology has assisted in many ways but for experimental work a physical presence at experimental site is essential and expected. LHC experiment collaborators are required an entry fee for a new institution joining an experiment and than an annual  membership fee ~\cite{Ref11,Ref12,Ref14}. If one has a federal grant it covers these fees automatically. The fee enables one to be a default author on all publications from these experiment.

\subsection{Institutional culture}

There are many challenges for pursing research as a faculty at a PUI due to the very nature of its setup as an institution and this has been described in details ~\cite{Ref1b1,Ref2,Ref3,Ref4,Ref5,Ref6,Ref7}. While HEP research has many commonalities with other fields of research it does have its own very specific challenges and they add a layer of its flavor to all common challenges with other areas of research. 

\subsubsection{Focus of student research}
Both theoretical and experimental HEP deal with fundamentals of  nature and have years of preparation time in terms of studying physics and math to be able to contribute at the level of research. In other words, there is big gap of time between start of learning to the start of research. Experimental HEP has an advantage that it involves lots of instrumentation and students as young as undergrads can contribute to it in some form of labor or small software projects. In other words, involving  undergrads and Master students in research at a PUI is seen mostly as a teaching process and less of them giving back to the faculty in research. A direct help like a PhD student or postdoc would do, is not easy or expected. This places the faculty in a very difficult situation where on one side one has to perform at the level of R1 and lab collaborators but with minimal useful help locally at PUI. However, it is still experience for undergrads and Master students as it prepares a STEM workforce in the due process and federal agencies like NSF do give a great value to these efforts. In fact NSF-RUI research proposals solicitation requires an impact statement  which points essentially to the difference the research makes for these students. 

Background education through coursework can be a challenge for students attending a PUI, especially those PUIs with a small physics department or where physics is part of a larger umbrella department such as Natural Sciences. Physics groups at PUIs often carry a disproportionately large amount of service burden of preparing non-physics majors compared to those at R1 institutions.

%\subsubsection{Student support -- ID the goal for this heading?}
%Higher level courses

\subsubsection{Teaching responsibilities}

%\textit{Written by Bhujyo: please feel free to edit as necessary}

The additional teaching and often service-teaching burden that faculty at PUIs face, has several undesired consequences. First, physics faculty at PUIs have a disproportionately large teaching burden. Whereas at an R1 institution faculty may have the responsibility of performing research with up to 60\% of their time, at a typical PUI faculty are limited to roughly half that. Second, faculty at smaller institutions often have to devote an unfairly large amount of time toward preparing and teaching non-particle physics courses. This draws their attention away from both particle physics research, as well as advanced courses that would be considered bread-and-butter for a student of particle physics, such as special relativity and quantum physics. Third, smaller institutions that have established engineering and professional degrees appear to recruit and retain only a handful of physics majors every year. As such, often upper-level physics courses at smaller institutions are either underenrolled or can only be offered at lengthy intervals by creating a larger enrollment pool from students spanning different years of college experience. Last but not least, PUIs often lack the resources to offer an advanced version or part II of upper-level classes that are, on the other hand, regularly available to students at R1 institutions.

The above challenges have a direct and adverse impact on student access to particle physics, often in the form of a lack of background preparation. This is especially acute for students interested in particle theory research, as many of the theoretical and computational tools necessary for graduate-level research in particle theory are taught in advanced physics and mathematics courses. The absence of these courses in the required undergraduate curriculum leaves the average physics major at a PUI unexposed to some of the fundamental tools necessary for continued success in pursuing a career in particle physics research.

Despite the above challenges to particle physics training and education, students from PUIs often pursue successful research careers in particle physics. In large part, the success of students who attend PUIs can be attributed to their enhanced interaction with faculty members when compared with the students attending R1 institutions.

\subsubsection{Research support}

Another significant challenge for HEP research at PUIs, particularly experimental HEP research, is lack of institutional support for research programs with significant external collaboration. Most research within a physics department at a PUI happens in local labs with small-scale equipment purchased and maintained by the PI. But large HEP experiments are hosted at national or international laboratories, and their operation and maintenance is funded collectively by the different countries involved in the project. The concept of an ``authorship fee" (or even a collective "author list"!) is difficult to understand by colleagues and administrators who are not familiar with HEP research structures. New faculty at PUIs will likely face more difficulty communicating research needs to their institutional leaders than new faculty at non-PUIs, such as the need for time to prepare a strong funding proposal during the first year of teaching. 

Similarly, access to host laboratories for HEP experiments often requires legal agreements between the laboratory and the home institution of each researcher. Faculty at PUIs typically cannot benefit from long-standing agreements between a well-established HEP group and a laboratory and have the extra hurdle of explaining to university administration and legal teams why these agreements or contracts are important for their research.

\subsection{Specific challenges for CCs} 

Community colleges are affordable and typically have more open admission policies than 4-year institutions. Approximately 40\% of undergraduate students in the United States are currently enrolled in CC, and a large proportion of those students are from demographics typically underrepresented in STEM fields \cite{ref15,ref16}. CC is a great place for students to start their path towards higher education. Nearly 80\% of CC students indicate their goal is to earn a bachelor’s degree \cite{ref16}. However, CC students face many challenges that unfortunately lead to nearly 70\% of students dropping out before completing their degree \cite{ref17}.    

CC students don’t always take the same traditional path that many students at 4-year institutions take, graduating high school then going on to start at a 4-year college. CC students are often returning to school after time off or starting at a nontraditional age. Students who fall into these categories are likely to have other responsibilities, such as family commitments or work, that put them at a higher likelihood of leaving school \cite{ref18}. Just over 60\% of full-time students and 72\% of part-time students work at the same time they’re enrolled \cite{ref15}. Another challenge for many CC students is that they haven’t received sufficient academic preparation \cite{ref18}. Without preparation, students can quickly fall behind, become discouraged, and withdraw from the CC.  

Having access to and participating in STEM research increases a CC students' chances of working toward a higher degree \cite{ref21}. This poses a particularly difficult challenge because research tends to happen outside normal class hours. The extra time commitment needed for research reduces the number of CC students who are available to participate due to their responsibilities outside school.

%\section{Goals for change}
\section{Paths toward broader research access}

\begin{figure}
    \centering
    \includegraphics[width=1.0\textwidth]{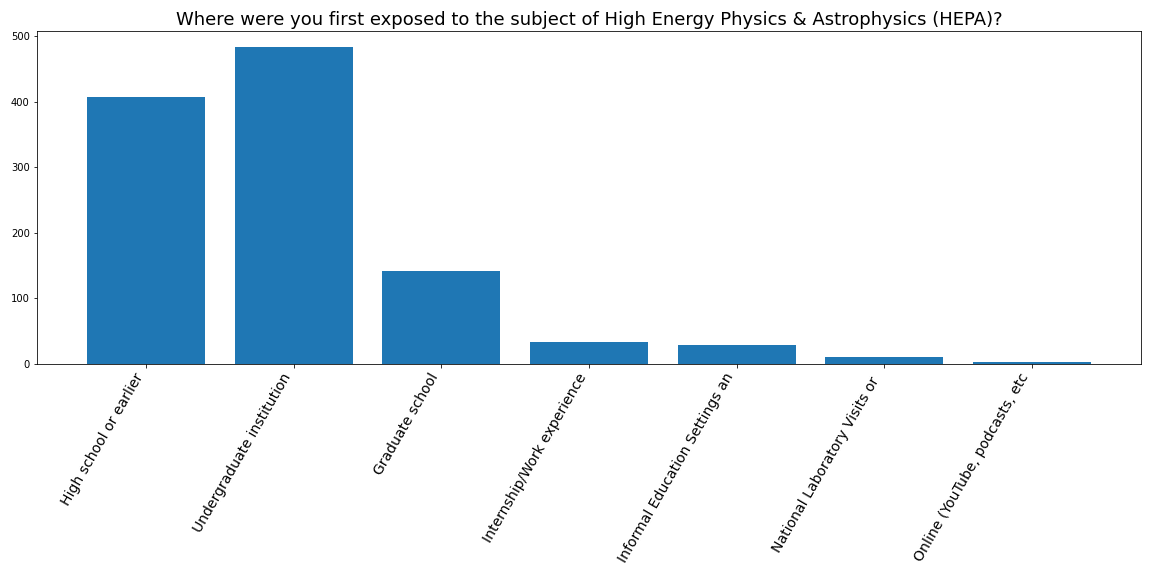}
    \caption{Question 8 from the survey. "Where were you first exposed to the subject of High Energy Physics and Astrophysics (HEPA)}
    \label{fig:q8}
\end{figure}

There are several actions that funding agencies, collaborations, and the HEP community can take to expand research opportunities at PUIs, which have an important role to play in preparing students and providing new career avenues within HEP. Students and postdocs in HEP would have broader career options within academia if HEP research at PUIs becomes more accessible, given the limited number of faculty positions available each year at non-PUIs. Undergraduate students at PUIs would also gain more opportunities to learn about particle physics and engage in early research, which is increasingly important for admission to graduate programs. Figure~\ref{fig:q8} shows that most people responding to the Snowmass Early Career community survey were first exposed to particle physics during their undergraduate education. Particle physics faculty at PUIs can bring this exposure to a new group of students, potentially more diverse than their non-PUI peers, that have chosen to pursue undergraduate education at these smaller institutions. Since PUIs offer tremendous potential to tap a diverse pool of candidates for the future STEM work force, it is important that HEP community and funding agencies strengthens participation by PUIs in HEP.

Expanded access to research opportunities would also benefit CC students. Data indicate that a large percentage of CC students want to excel in CC and move on to earn bachelor's degrees, but only a small percentage of students are able to for a number of reasons mentioned in section 2.4. Participating in research increases a CC student's chances of success. Research opportunities need to be made accessible to CC students who juggle many responsibilities during their enrollment and don't have a significant amount of time to dedicate solely to research.

%\textit{Julie will write a draft of this section 2/21 after the meeting}

%Goal statement, why should the problem be addressed. These are likely 1-2 sentences each, similar to short "Problem statement" section 
%\begin{itemize}
%    \item Specific goals / rationale for PUI  (ALL state desired, merge them)
%    \item Connection to goals for DEI  (Julie on perspective from CMS DEI)
%    \item Highlight UG as most common entry point to HEP, shown in graph from survey.
%    \item Goal of broader hiring paths for HEP postdocs, with better training for PUI applications
%    \item Specific goals / rationale for CC (Kathrine L): Data indicate that a large percentage of CC students want to excel in CC and move on to earn bachelor's degrees, but only a small percentage of students are able to for a number of reasons mentioned in section 3.4. Participating in research increases a CC students' chances of success. Research opportunities need to be made accessible to CC students who juggle many responsibilities during their enrollment and don't have a significant amount of time to dedicate solely to research.
%\end{itemize}

The challenges PUIs face when engaging in HEP research could be alleviated if collaborations, funding agencies, and the HEP community followed one or more of the following paths toward broader research access. 

\subsection{Research funding}

Two changes to research funding opportunities would significantly improve access to HEP research at PUIs: more opportunities from the Department of Energy, and increasing flexibility for connections between faculty at PUIs and nearby non-PUIs with HEP research groups. 

Particularly for participation in large experimental collaborations, faculty members and students must have a home institution that is ``known" to the experiment. Even if a faculty member at a PUI acquires research funding, perhaps through a program such as the National Science Foundation's ``Research at Undergraduate Institutions", unless their institution is recognized as a member of the experiment participation is not possible. Affiliation with a nearby non-PUI HEP research group can help solve this access problem. 

Affiliation with a non-PUI group also makes it more likely that a PUI faculty member will be able to acquire funding, and begin or continue research in the experiment while a proposal is reviewed. Such a connection provides something of a ``safety net" for the PUI faculty member's proposal, since they can collaborate on projects with graduate students and postdocs at the non-PUI. This affiliation can take the form of a ``visiting scientist" appointment, an adjunct faculty appointment, or similar. Access to the experiment, such as computing accounts, come through the non-PUI group. 

\begin{figure}[ht]
\centering
\includegraphics[width=\textwidth]{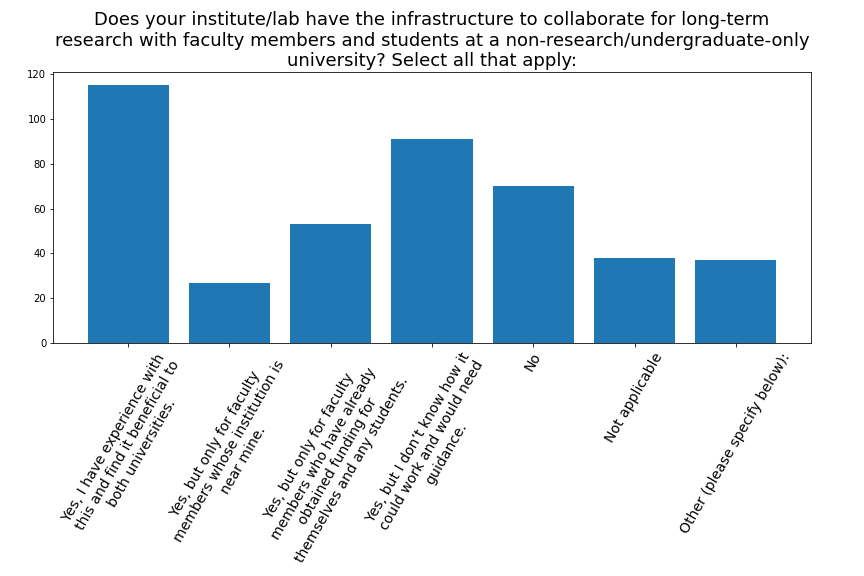}
\label{fig:R1}
\caption{Survey data on connections between PUIs and non-PUIs.}
\end{figure}

The non-PUI group benefits from an additional mentor for students, and may build connections with potential incoming graduate students. Cross-pollination of ideas and expertise between the PUI and non-PUI groups can improve the creativity and scientific impact of both, particularly by retaining talented researchers at PUIs that would otherwise leave the field. The SEC community survey queried whether or not non-PUI faculty members felt their institutions could participate in such an affiliation. The results, shown in Fig.~\ref{fig:R1}, suggest that many non-PUI groups are open to helping improve research access for their PUI peers, through several need examples or guidance on how to set up such an affiliation.

Affiliations between a PUI faculty member and a non-PUI HEP group can also provide the PUI faculty member with authorship in the experimental collaboration, as long as the PUI faculty member has been awarded research funding from a national agency. This is simplest when the PUI faculty member and the non-PUI group are funded through the same agency, such as the National Science Foundation. When the non-PUI group is funded through the U.S.~Dept.~of Energy and the PUI faculty member is not, extra manual intervention is required each year to properly account for the PUI faculty member when the agencies contribute per-author fees to the experiment. If the DOE were to provide a funding avenue for PUIs, this would effectively double the number of non-PUI groups with whom a PUI researcher could easily affiliate. Funding PUI research would be a novel way to improve the agency's ability to ``create a vibrant scientific ecosystem''~\cite{DOEweb} in the United States. In what follows we provide a list of programs that are of value and should be considered for continued funding by federal agencies to increase research participation at PUIs and CCs.

Faculty and students at PUIs can also participate in research taking place at non-PUIs and national labs via programs that provide summer research opportunities for those without other sources of funding. One example is the DOE Visiting Faculty Program (VFP). The Visiting Faculty Program (VFP) is a DOE-sponsored program that aims to increase the research competitiveness of faculty at institutions typically underrepresented in the research community (non-R1/R2, unless URM campus) \cite{ref19}. For 10 weeks during the summer, a faculty member and up to two graduate or undergraduate students collaborate with a scientist/engineer at one of the DOE national labs on a research project of mutual interest \cite{ref19}. 
    
The VFP offers many benefits to faculty, students, and laboratory staff. Participating faculty members and undergraduate students receive stipends during the 10-week period. In addition, the host laboratory covers the cost of housing and travel to and from the lab. The team can continue work on their research project for up to three consecutive summers through the VFP. This is particularly helpful for long term projects that can't be completed in one summer. The VFP fosters collaboration and builds lasting relationships between DOE laboratory staff and faculty. Both faculty and participating students benefit from exposure to state-of-the-art technology and equipment at lab facilities, features they likely don't have access to at their home institutions. Students are able to network with lab scientists/engineers, learn valuable professional development skills, and gain hands-on research. The VFP reaches a broad group of faculty and students that are underrepresented in the research community. This is crucial to expanding the diverse recruiting network and helping attract the future workforce for the national labs.

Another example is internship programs that connect summer students to the major experiments. In
2021, member of the CMS experiment began an effort to engage students at non-CMS institutions
in upgrade projects funded as part of the MREFC effort. This led to the CMS MREFC-EPO 
Underrepresented Minority Internship, scheduled to run in Summer 2022. Five (5) students, all of
whom are Black or Hispanic, will work with groups at Boston University, Purdue University, 
and University of Kansas. This experience would not have happened at their home institutions, and
while they have the opportunities to apply to traditional REUs, this type of targeted program 
is often necessary to reach students who might not see themselves engaging in this research. 
The CMS could absolutely have run this effort without any PUI involvement, but as was mentioned
before, the effort is spearheaded by two co-conveners who are faculty at PUIs (Bellis, Hogan) and
have brought a particular perspective and emphasis to the work, making it a priority. 

Similarly, faculty at PUIs and CCs who are part of a larger effort, whether theory or experiment, are
often plugged in to opportunities in the broader community. 
For example, IRIS-HEP\footnote{\url{https://iris-hep.org/}}
and DIANA-HEP\footnote{\url{https://diana-hep.org/}} are two NSF-funded projects aimed at
improving the software tools used in the HEP community. Both have sponsored undergraduate fellowships
available to anyone in the country. A well-motivated undergrad with a working knowledge of Google
can seek out fellowships like these, but the reality is often quite different. Undergrads can 
be overwhelmed by the options available to them and can have an event horizon that ends
at the borders of their campus. It takes a guiding mentor to point them in the right direction and 
this is where a faculty mentor can have an outsized impact, even if it is just passing along 
the right opportunties to students. But if the faculty member is not connected to the broader
community efforts through collaborative research, it is hard to stay aware of these 
opportunities. 

Finally, grantsmanship is a key element that determines the success of a research program. Yet, often little grantsmanship training and support is currently available to a large section of the HEP community -- graduate students, postdoctoral researchers, and junior faculty members at PUIs. Federal funding for programs that support grant-writing workshops will be able to bridge this gap and increase participation in HEP research. Under such programs, researchers who have been successful in securing grants will be paired with students and early-career professionals, creating a network of professionals invested in broadening access to HEP grants. Such a program can be modelled after the Broadening Experiences in Scientific Training (BEST) initiative of the National Institutes of Health (NIH) -- a program originally developed to strengthen the biomedical research workforce \cite{NIHBEST}.

\subsection{Large experiment participation}

Large experimental collaborations could adopt various methods of expanding access for small groups of researchers like those at PUIs. The primary need is for membership in collaborations to be accessible for small institutions with limited funding. The governing bodies of large experiments could:
\begin{itemize}
    \item Reevaluate large fixed ``entry fees" per institution, if they exist. Such fees place a disproportionate burden on PUIs that lack resources compared to non-PUIs. An entry fee structure that scales based on the number of Ph.D.~holders in the institution would be more equitable. One option might be a ``single PI" membership. 
    \item Consider implementing ``light" membership forms that are low cost but not time limited. Faculty members at PUIs likely care more about access to the experiment and laboratory facilities for themselves and their undergraduate students than full authorship on all collaboration papers. Existing membership forms that do not incur costs to the institution often have a time limitation which makes them ill-suited to long-term collaboration. 
    \item Provide financial support mechanisms for PIs to travel to important full-experiment meetings. Faculty members at PUIs often have much less travel support than non-PUI peers. 
    \item Continue to improve options for remote participation in service tasks, especially operational shift work. When full membership in a collaboration requires a certain quota of shift work per institution, small groups at PUIs are at a significant disadvantage because travel to a host laboratory is typically not possible during the academic year.
\end{itemize}

%Idea of flipped VFP with a scientist visiting a university/college? “Lighter” program without pay help, just facilitating/organizing projects in need of help/input from lab scientist? OR perhaps provides some student support along with lab scientist presence? Could propose such an idea if it fills a gap.

\subsection{Institutional culture}
%\textit{Ben and Matt}

Because of the nature of PUIs and liberal arts colleges where teaching and a student-centered
experience is emphasized, faculty must work hard to find time to do research and to be
very efficient in that research. The trade-off is that they are regularly engaging with 
best practices in pedagogy and mentoring. There are some specific actions that the community 
and funding agencies could take which would have an outsized impact on PUI faculty. 

\begin{itemize}
    \item Currently the practice of using grant funds to buy out of teaching duties is 
    generally discouraged by NSF and DOE. While this is perhaps understandable for faculty 
    at research institutions, at PUIs, faculty are often teaching 3 or 4 course per semester. Allowing
    faculty to request funds to support reassigned time, for example in NSF RUI (Research at
    Undergraduate Institutions) would have a huge impact. 
    \item While the situation is improving and we acknowledge those collaborators who are 
    particularly supportive, the community would benefit from a more global shift in the perception and acknowledgement that undergraduate research experiences are {\it key} to engaging a broader
    section of the student population. Related, an awareness that PUI faculty have much to offer
    experiment-wide training and educational activities.
    \item Because faculty at PUIs are often incentivized to stay current with educational research and
    best-practices, they often have something to offer their collaborators, particularly in 
    helping train graduate students and post-docs to be better teachers. This could help those
    individuals be better prepared for the job market.
    One suggestion would be to find opportunities for grad students or post-docs to teach at formal collaboration events (e.g. CMS Data Analysis School) and also find ways to provide constructive feedback to help them improve their teaching.
    \item The authors noticed that the Snowmass Climate Survey did not ask about ``Teaching" as part of the desired skills that one might acquire as part of their HEP education. This points to a potential blind spot by the community or that ``learning how to teach" is not an important skill. A broader community-wide shift in valuing good teaching would be welcome.  
    \item PUI administrators may not always appreciate the benefit of having faculty participate
    in research activities with large, external collaborations, even when those faculty are funded. Coordinated communication from experimental leaders that is directed to PUI administrators, 
    extolling the features of high energy physics research alongside highlighted participation 
    would go a long way to contributing to tenure and promotion for PUI faculty. %ddm - many administrators see research participation by faculty a distraction from teaching in a context of tuition driven enrollment goals
    \item Developing a sense of community among the PUIs and CCs who are doing this work would be helpful. This could manifest as special session at APS or DNP or similar meetings, or even dedicated workshops for faculty members to share experiences and brainstorm on how to make the most of their connections. 
\end{itemize}

\subsection{Specific ideas for community colleges}

Programs such as the Visiting Faculty Program (VFP) and Community College Internships (CCI) provide significant benefits to the CC faculty and students who participate in them. These programs are specifically designed for faculty and students at institutions, such as community colleges, that receive much less research funding than R1/R2 universities. The VFP and CCI can help strengthen relationships between CC and DOE national labs which in turn, could increase the number of CC that join large collaborations with national labs and other research institutions. This would be highly beneficial to CC students and faculty as well as laboratory staff. 

The CCI is an internship program specifically for CC students. Through the program, students hone their technical skills and participate in research projects at DOE national labs. During the internship, students work closely with lab scientists and engineers. One key benefit of this program is that there are flexible internship dates to accommodate the needs of CC students juggling many commitments. The CCI has 10-week internship terms over the summer, fall, and spring. During the spring and fall terms, students can opt to follow a "flex-schedule" that is adaptable to meet their availability \cite{ref20}. Another perk of the CCI is that the students receive a stipend for the 10-week term. This means that students won't have to choose between a job and research experience, the CCI provides both. 

During the CCI, students work closely with lab staff, building relationships that could lead to additional internships and potentially future full-time positions. Students also have the opportunity to participate in professional development workshops, lab tours, and scientific lectures/seminars \cite{ref20}. As studies have shown, CC students significantly benefit from participating in research during their time in school \cite{ref21}. The CCI program has had an average of 78 students participate each year from 2014 - 2020 \cite{ref20}. While it is fantastic to see this level of engagement, this program should be more widely advertised, where possible, so a larger number of students can benefit from the valuable experience the CCI provides.

%\begin{itemize}
%    \item Resources to make research experiences financially equivalent to other jobs during school?
%\end{itemize}

\section{Examples of PUI research programs}

Bhujyo Bhattacharya leads a theoretical particle physics research group at Lawrence Technological University (LTU) in Southfield, Michigan. With research funding from the National Science Foundation through an RUI grant, Bhattacharya engages undergraduate students in particle physics research projects. At a small institution like LTU, the biggest challenges revolve around the availability of resources. Advanced physics classes, such as quantum mecahnics and electromagnetism, are not offered on a regular basis. Furthermore, due to lack of interest (low enrollment), only one-semester versions of these courses are offered. This makes it challenging for the students who work with Bhattacharya to gain enough background knowledge to begin research. Bhattacharya overcomes these challenges by introducing students to research even in his first year classroom. LTU's College of Arts and Sciences has received funding for Inclusive Excellence from the Howard Hughes Medical Institute (HHMI). As a senior personnel in this grant, Bhattacharya uses funds from this grant to create Course-based Research Experience (CRE) project for his first-year introductory-physics students. Students working with Bhattacharya create ``Computational Essays'' to demonstrate a physics principle. This process provides students training in computational methods -- they develop coding skills that prove to be extremely beneficial for preparing students for NSF-funded research projects in theoretical particle physics.

Julie Hogan is an Assistant Professor of Physics at Bethel University in Saint Paul, Minnesota. She and a cohort of undergraduate research students participate in the Compact Muon Solenoid (CMS) experiment. Hogan's research program was launched out of postdoctoral work at Brown University, and she remains a CMS author via affiliation with Brown. Following an access crisis while attempting to bring students onto the CERN site in summer 2018, Bethel University has become a "Cooperating Institute" in CMS, which allows Bethel students to register properly as CERN users and to be listed under Bethel University on any papers. Through support from the NSF RUI program, Hogan's students have traveled to Fermilab and CERN to work on upgrades to the CMS silicon strip tracker for the High-Luminosity LHC and searches for vector-like quarks using deep neural networks. The group's favorite research activity is rendering new CMS simulation events into LEGO bricks for a "particle flow" algorithm tutorial offered at the Fermilab LHC Physics Center during (pre-pandemic) summers. The students have presented their work at conferences, earned credit as authors on vector-like quark search papers, and several have now gone on to graduate study in physics, math, or data science. Bethel's challenge over the next two years, when their Cooperating Institute membership will expire, is to navigate a more sustainable method of continuing CMS research into the future. 

Matt Bellis is an Associate Professor in the Department of Physics and Astronomy at Siena College, 
located just outside of Albany, NY. Bellis and the Siena students are part of the CMS collaboration
through a collaboration with Cornell University and are listed as Cornell members in the CMS
database. This is a limited membership and Bellis and his students only receive authorship on
papers to which they have directly contributed, though other options are being explored. 
Bellis is supported through an NSF RUI grant and engages students in research on rare decays
of the top quark, supporting the CMS open data effort, and novel approaches to education and outreach. 
The work and experiences from this research have filtered into the classroom in the form 
of an introductory programming course for all freshman physics majors, an upper-level nuclear
and particle physics course, and a course designed for non-science majors: {\it Quarks, Quanta, and
Quasars}. Siena has participated in CMS research since 2013 and more than 20 students have been
supported by NSF grants on this work. While some have gone on to grad school for particle physics
(with 1 PhD so far and 1 to finish in 2022), the majority have gone on to grad school or jobs
in medical physics, data science, civil engineering, mechanical engineering, atmospheric science
and more, proving that particle physics research is a superb training ground for a wide variety
of career goals. 

\section{Conclusion}

Employment in HEP fields at R1 Universities and National Labs is quite saturated. In order to foster a more vibrant and diverse HEP community, opportunities and support for undergraduate research need to be expanded. One way to accomplish this goal is through improved inclusion and integration of PUIs and CCs in the HEP community. These undergraduate-focused programs typically cater to a different cross-section of the student population and with their focus on education, are a wonderful compliment to the research-driven universities and labs.

PUIs and CCs face a variety of challenges that inhibit their ability to more fully participate in and contribute to the HEP community. These include, but are not limited to a lack of infrastructure or support system, a lack of funding, and large teaching loads. However, many of these challenges can be mitigated with some creative collaboration, additional training, and adjustments to funding. 

Funding agencies could have a massive impact on enhancing HEP research at PUIs and CCs by expanding or creating additional funding opportunities for these programs. Large collaborations could consider ways to mitigate barriers prohibiting more full engagement with PUIs and CCs. The HEP community as a whole could better recognize and advocate for the value that PUIs and CCs bring to the community. 

In short, the HEP community would benefit greatly from more fully engaging with and supporting PUIs and CCs.

\subsection*{Acknowledgements}
Contributions are based upon work supported by the National Science Foundation under Grant No.~PHY-1913923 (MB), PHY-2013984 (BB), and PHY-1806415 and PHY-2110972 (JH).

\bibliographystyle{utphysmod}
\typeout{} 
\bibliography{main.bib}

%time = (range(0,20,1))

\end{document}

%% file: workshopsymbols.tex
\def\beq{\begin{equation}}
\def\eeq#1{\label{#1}\end{equation}}
\def\eeqn{\end{equation}}

\newenvironment{Eqnarray}%
   {\arraycolsep 0.14em\begin{eqnarray}}{\end{eqnarray}}
\def\beqa{\begin{Eqnarray}}
\def\eeqa#1{\label{#1}\end{Eqnarray}}
\def\eeqan{\end{Eqnarray}}

\let\bar=\overbar

\def\lsim{\mathrel{\raise.3ex\hbox{$<$\kern-.75em\lower1ex\hbox{$\sim$}}}}
\def\gsim{\mathrel{\raise.3ex\hbox{$>$\kern-.75em\lower1ex\hbox{$\sim$}}}}

\def\del{\partial}
\def\Dslash{\not{\hbox{\kern-4pt $D$}}}
\def\dslash{\not{\hbox{\kern-2pt $\del$}}}
\def\pslash{\not{\hbox{\kern-2pt $p$}}}
\def\ETmiss{\not{\hbox{\kern-4pt $E$}}_T}

\def\Dlr{\mathrel{\raise1.5ex\hbox{$\leftrightarrow$\kern-1em\lower1.5ex\hbox{$D$}}}}

\def\MSB{{\bar{M \kern -2pt S}}}
\def\msb{{\bar{\scriptsize M \kern -1pt S}}}

\def\drb{{\bar{\scriptsize D \kern -1pt R}}}